\documentclass[conference]{IEEEtran}

\IEEEoverridecommandlockouts
\usepackage{graphicx}
\usepackage{amsmath,mathtools}
\usepackage{amssymb}
\usepackage{bbm}
\usepackage{xcolor}
\usepackage{tabularx}
\usepackage{algorithm}
\usepackage{algpseudocode}
\usepackage{tikz}
\usetikzlibrary{arrows.meta, positioning, shapes.multipart}
\usepackage{geometry}
\usetikzlibrary{decorations.pathreplacing}
\def\BibTeX{{\rm B\kern-.05em{\sc i\kern-.025em b}\kern-.08em
    T\kern-.1667em\lower.7ex\hbox{E}\kern-.125emX}}
\usepackage{url}
\usepackage{cite}
\newcommand{\llrrparen}[1]{
\left(\mkern-3mu\left(#1\right)\mkern-3mu\right)}
\begin{document}

\title{Waveform-domain NOMA: An Enabler for ISAC in Uplink Transmission}
\author{\IEEEauthorblockN{Jialiang~Zhu\textsuperscript{*},\;
Hamza~Haif\textsuperscript{$\dagger$},\;
Abdelali~Arous\textsuperscript{$\dagger$},\;
Hüseyin~Arslan\textsuperscript{$\dagger$},\;
Arman~Farhang\textsuperscript{*}
} 
\IEEEauthorblockA{\textsuperscript{*} Department of Electronic \& Electrical Engineering, Trinity College Dublin, Ireland \\
\textsuperscript{$\dagger$} Department of Electrical and Electronics Engineering, Istanbul Medipol University, Turkey
\\Email: zhuj3@tcd.ie, \{hamza.haif, abdelali.arous\}@std.medipol.edu.tr, huseyinarslan@medipol.edu.tr, arman.farhang@tcd.ie}
\vspace{-1 cm}
\thanks{
 This publication has emanated from research conducted with the financial support of Research Ireland under Grant number 18/CRT/6222 and by the Scientific and Technological Research Council of Türkiye (TUBITAK) under the Grant Number 123E514. For the purpose of Open Access, the author has applied a CC BY public copyright licence to any Author Accepted Manuscript version arising from this submission.
}
}


\maketitle

\begin{abstract}
According to the recent 3GPP decisions on 6G air interface, orthogonal frequency-division multiplexing (OFDM)-based waveforms are the primary candidates for future integrated sensing and communication (ISAC) systems. In this paper, we consider a monostatic sensing scenario in which OFDM is used for the downlink and its reflected echo signal is used for sensing. OFDM and discrete Fourier transform-spread OFDM (DFT-s-OFDM) are the options for uplink transmission. 
When OFDM is used in the uplink, the power difference between this signal and the echo signal leads to a power-domain non-orthogonal multiple access (PD-NOMA) scenario.
In contrast, adopting DFT-s-OFDM as uplink signal enables a waveform-domain NOMA (WD-NOMA). 
Affine frequency-division multiplexing (AFDM) and orthogonal time frequency space (OTFS) have been proven to be DFT-s-OFDM based waveforms. This work focuses on such a WD-NOMA system, where AFDM or OTFS is used as uplink waveform and OFDM is employed for downlink transmission and sensing.
We show that the OFDM signal exhibits additive white Gaussian noise (AWGN)-like behavior in the affine domain, allowing it to be modeled as white noise in uplink symbol detection. To enable accurate data detection performance, an AFDM frame design and a noise power estimation (NPE) method are developed. Furthermore, a two-dimensional orthogonal matching pursuit (2D-OMP) algorithm is applied for sensing by iteratively identifying delay–Doppler components of each target. Simulation results demonstrate that the WD-NOMA ISAC system, employing either AFDM or OTFS, outperforms the PD-NOMA ISAC system that uses only the OFDM waveform in terms of bit error rate (BER) performance. Furthermore, the proposed NPE method yields additional improvements in BER.
\end{abstract}

\begin{IEEEkeywords}
AFDM, OTFS, ISAC, WD-NOMA, DFT-s-OFDM
\end{IEEEkeywords}

\vspace{-0.4cm}
\section{Introduction}
Integrated sensing and communication (ISAC) is considered as a key technology for future wireless systems. It supports not only ultra-reliable and high-capacity data transmission but also advanced functionalities such as environmental awareness, localization, and monitoring\cite{isac}. These functionalities are particularly critical for emerging applications including autonomous driving, smart manufacturing and unmanned aerial systems. The main challenge in realizing ISAC lies in the design of waveforms that jointly optimize sensing and communication performance, using single or distinct waveforms \cite{isacwaveform}. 
Although single-waveform solutions achieve high spectral efficiency, they typically require compromises that limit either sensing accuracy or communication reliability \cite{highrange}. 
Such issues can be avoided by using distinct waveforms for sensing and communications in time division duplex (TDD) mode, which comes at the expense of lower spectral efficiency \cite{isactdd}.
To address this shortcoming, power-domain non-orthogonal multiple access (PD-NOMA) was proposed in \cite{pdnoma1}, where communication and sensing signals using the same waveform are superimposed. However, PD-NOMA requires successive interference cancellation (SIC) which is prone to error propagation and channel estimation errors \cite{pdnoma}. As an alternative approach, waveform-domain (WD)-NOMA superimposes sensing and communication signals that are in distinct transform domains to alleviate the interference among them \cite{wdnoma}.

Well-established 4G/5G waveforms, orthogonal frequency-division multiplexing (OFDM) and discrete Fourier transform spread OFDM (DFT-s-OFDM), along with certain enhancements are being considered by 3GPP as the air interface technologies for 6G \cite{3gpp}. This makes OFDM and DFT-s-OFDM the primary waveform candidates for future ISAC systems. In this paper, we consider a monostatic sensing scenario where OFDM is used for downlink transmission, and its reflected echoes are used for sensing. When OFDM is also used in the uplink, the power difference between the uplink signal and the reflected echo results in a PD-NOMA scenario at the base station (BS). From a communication perspective, the waveform choice must cope with the high environment mobility. OFDM is known to have a poor performance under time-varying channels \cite{ofdm}. This limits its communication reliability in ISAC scenarios.
Alternatively, using DFT-s-OFDM for the uplink leads to a WD-NOMA.
Waveforms such as affine frequency-division multiplexing (AFDM) and orthogonal time frequency space (OTFS), have been investigated recently  \cite{afdm, otfs}. These waveforms multiplex data symbols in domains that better capture the channel’s stationary properties, thereby improving robustness to mobility-induced inter-carrier interference (ICI). 
As shown in \cite{dft_afdm} and \cite{scfdma}, both AFDM and OTFS can be realized using DFT-s-OFDM with interleaved subcarrier allocation. Hence, AFDM and OTFS can be considered as DFT-s-OFDM based waveforms to form WD-NOMA.

Motivated by the above considerations, this paper investigates a WD-NOMA system for ISAC with high-mobility users. In this system, the BS transmits OFDM as a downlink signal for communication and receives its reflected echoes from surrounding targets for sensing. Meanwhile, user equipments (UEs) transmit DFT-s-OFDM based waveforms as the uplink signal, which is received by the BS. As the combination of OTFS and OFDM for WD-NOMA has been studied in \cite{otfsnoma}, this work focuses on AFDM as the uplink waveform.
The main contributions of this paper are summarized as:
\begin{itemize}
    \item Based on the DFT-s-OFDM based waveforms, we consider an ISAC system which uses OFDM for downlink, and AFDM or OTFS for uplink signaling. This ISAC system uses non-orthogonal, uncoordinated reception at the BS where both the uplink and the reflected downlink signal are fully overlapped in time and frequency.  
    \item Leveraging the waveform-domain properties, OFDM exhibits AWGN-like power distribution in the affine domain. We introduce a guard-aided noise power estimator (NPE) within the AFDM frame that uses a predetermined set of guard carriers to estimate the OFDM interference level. 
    \item Based on this estimation, the interfering OFDM echo signal in the affine domain can be canceled and the uplink AFDM symbols are detected using a minimum mean-square error (MMSE) detector. Following that, the uplink signal plus channel is then reconstructed and canceled in the time domain to obtain the pure the reflected OFDM echo and channel noise. Subsequently, two-dimensional orthogonal matching pursuit (2D-OMP) is applied to the post-cancellation signal for target parameter estimation.
\end{itemize}
Simulation results demonstrate that the WD-NOMA ISAC achieves better bit error rate (BER) performance compared to PD-NOMA ISAC system which uses only OFDM. Proposed NPE further improves BER. 

\textit{Notation:}
Scalars, vectors, and matrices are denoted by regular lowercase, bold lowercase and bold uppercase letters, respectively.
The superscripts $(\cdot)^{\rm{H}}$, $(\cdot)^{\rm{T}}$, $(\cdot)^{\rm{*}}$ and  $(\cdot)^{\rm{-1}}$ indicate Hermitian, transpose, conjugate and inverse operations, respectively.
The operators $\otimes$ denotes Kronecker product.
${\rm{diag}} \{ \mathbf{x} \}$ represents a diagonal matrix with the elements of the vector $\mathbf{x}$ on its main diagonal.
$\mathbb{E} \{ \cdot\}$ and $\mathrm{var} \{ \cdot \} $ denote the expected value and variance of a variable, respectively. 
$\mathbf{I}_{N}$ is the identity matrix of size $N$.
$\mathbf{{F}}_{N}$ is the normalized $N$-point DFT matrix with the elements $[\mathbf{F}_{N}]_{m,n} =\frac{1}{\sqrt{N}} e^{-j \frac{2 \pi mn}{N}}$ for $m,n=0,\ldots, N-1$.

\section{System Model}
\begin{figure}[t]
\centering
\includegraphics[scale=0.32]{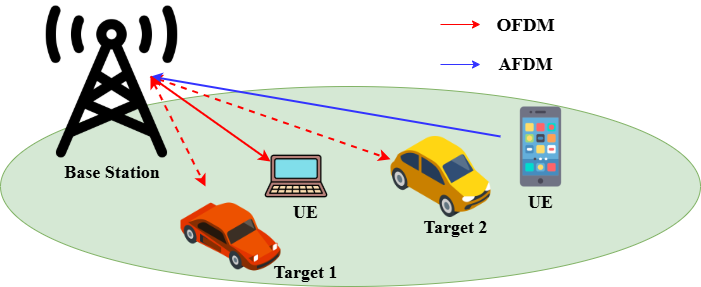}
\caption{Proposed ISAC system model. }
\label{fig:system_model}
\vspace{-0.4cm}
\end{figure}
We consider an ISAC system, shown in Fig.~\ref{fig:system_model}, where a single antenna BS performs monostatic sensing using the reflected downlink signal in the uplink phase while simultaneously serving an active UE in the uplink. In this work, we consider uplink and downlink signals use same number of subcarriers for simplicity but the number can be different in reality. OFDM is deployed for downlink transmission and occupies $N$ subcarriers. The transmit downlink data symbols, $x_\text{DL}[n]$, for $n=0,\ldots, N-1$, are chosen from a $\mathcal{M}$-QAM constellation with zero mean and unit variance. The data symbols are assumed to be independent and identically distributed (i.i.d.).
The discrete OFDM signal is represented by
\begin{equation}\label{eqn:ofdm_pulse}
\mathbf{s}_\text{DL} = \mathbf{A}_\text{cp}\mathbf{F}^\text{H}_{N} \mathbf{x}_\text{DL},
\end{equation}
where $\mathbf{x}_\text{DL}=[x_\text{DL}[0],\ldots,x_\text{DL}[N-1]]^\text{T}$,
$\mathbf{A}_{\rm cp} = [{\mathbf G}_{\rm cp}^{\rm T}, {\mathbf{I}_{N}^{\rm T}}]^{\rm T} $ is the CP addition matrix and the ${L}_{\rm cp} \times N$ matrix $\mathbf{G}_{\rm cp}$ includes the last ${L}_{\rm cp}$  rows of the identity matrix $\mathbf{I}_{N}$. 
Assuming line of sight (LoS) propagation between the BS and all sensing targets, the received OFDM echo signal at the BS, during the uplink phase while ignoring noise, can be written as 
\begin{equation}\label{eqn:r_sense}
\mathbf{r}_{\text{DL}} = \mathbf{H}_\text{DL} \mathbf{s}_{\text{DL}}, 
 \end{equation}
where $\mathbf{H}_\text{DL}=\sum_{p=1}^{P} h_p \, {\boldsymbol \Pi}^{\tau_p} \, {\boldsymbol\Delta}^{\nu_p}$ is the round-trip channel matrix and $h_p$ denotes the complex attenuation associated with the $p$-th target, determined by the two-way path loss of the target. Due to bidirectional propagation, comprising the forward path from the BS to the target and the return path from the target back to the BS, the delay and Doppler shifts are given by $\tau_p = \frac{2 r_p}{c}, \quad \nu_p = \frac{2 f_c v_p}{c},$ where $r_p$ and $v_p$ are the distance and the relative radial velocity between the $p$-th target and the BS, respectively, $c$ is the wave propagation speed and $f_c$ is the carrier frequency.
${\boldsymbol \Pi}$ is the unit cyclic permutation matrix, formed by circularly shifting the rows of the identity matrix, ${\mathbf{I}_{M+L_{\rm cp}}}$, by one position from top to bottom
and $\boldsymbol{\Delta}= \operatorname{diag} \{ [1,e^{-j \frac{2\pi}{M+L_\text{cp}}},\ldots,e^{-j \frac{2\pi(M+L_\text{cp}-1)}{M+L_\text{cp}}}]^{\rm T}\}$. 

In the uplink phase, we consider a user with $N$ subcarriers and the transmit data symbols $x_\text{UL}[n]$ for $n=0,\ldots, N-1$ that are chosen from a $\mathcal{M}$-QAM constellation with zero mean and unit variance. The data symbols are assumed to be independent and identically distributed (i.i.d.) and the uplink transmit signal is represented as 
\begin{equation}\label{eqn:comm}
\mathbf{s}_\text{UL} =  \mathbf{W}\mathbf{x}_\text{UL},
\end{equation}
where $\mathbf{x}_\text{UL}=[x_\text{UL}[0],\ldots,x_\text{UL}[N-1]]^\text{T}$ and
$\mathbf{W}$ is the modulation matrix for the uplink waveform. 
When OFDM is employed for uplink transmission, 
$\mathbf{W} \!\!\!=\!\!\! \mathbf{A}_{\text{cp}}\mathbf{F}_{N}^{\text{H}}$, 
and the system operates under PD-NOMA configuration. 
In contrast, when AFDM or OTFS is adopted for uplink transmission, 
the system functions under WD-NOMA configuration and the modulation matrices are denoted as 
$\mathbf{W}\!\!\!\!=\!\!\!\!\mathbf{A}_\text{cpp}\boldsymbol{\Lambda}_{c_1}^\text{H} \mathbf{F}^\text{H}_{N} \boldsymbol{\Lambda}_{c_2}^\text{H}$ and $\mathbf{W}\!\!\!\!=\!\!\!\!\mathbf{A}_\text{cp}(\mathbf{F}_{N_2}^{\rm H}\!\otimes\! \mathbf{I}_{N_1})$, respectively. $\boldsymbol{\Lambda}_{c} = \mathrm{diag}\{[1,e^{-j 2\pi c},\ldots,e^{-j 2\pi c (N-1)^{2}}]^{\rm T} \}$ and $c_1$ and $c_2$ are discrete affine Fourier transform (DAFT) parameters. The chirp-periodic prefix (CPP) addition matrix is given by $\mathbf{A}_\text{cpp}\!\!\!\!=\!\!\!\! [\mathbf{\Omega}{\mathbf G}_{\rm cpp}^{\rm T}, {\mathbf{I}_{N}^{\rm T}}]^{\rm T} $ where $\mathbf{\Omega}={\rm{diag}} \{[ e^{-j2\pi c_1(N^2-2NL_\text{cpp})},\ldots, e^{-j2\pi c_1(N^2-2N)} ]^{\rm T}\}$ and $L_\text{cpp}$ is the CPP length. Here, the ${L}_{\rm cpp} \times N$ matrix $\mathbf{G}_{\rm cpp}$ is constructed from the last ${L}_{\rm cpp}$ rows of the identity matrix $\mathbf{I}_{N}$. To avoid inter-symbol interference, $L_\text{cpp}$ must be larger than the channel delay spread. $N_1$ and $N_2$ are number of delay and Doppler bins for an OTFS frame respectively, which satisfy $N=N_1N_2$. 
The uplink signal propagates through a linear time-varying (LTV) channel.  
The received signal at the BS, ignoring noise, can be represented as  
\begin{equation}\label{eqn:r_comm}
\mathbf{r}_{\text{UL}} = \mathbf{H}_\text{UL} \mathbf{s}_{\text{UL}},
 \end{equation}
where $\mathbf{H}_\text{UL}$ is the uplink communication channel matrix in the time domain with the elements $[\mathbf{H}_\text{UL}]_{i,j}=h_\text{UL}[i-j,i]$ for $i,j=0,\ldots,K-1$ and $K=N+L_{\rm{cpp}}(L_{\rm{cp}})$.
Furthermore, $h_\text{UL}[j,i]$ is the uplink communication channel gain at the delay tap $j$ and sample $i$. 

The overall received signal at the BS consists of the uplink signal, the echo signal from the downlink phase and additive noise which is represented as 
\begin{equation}\label{eqn:r}
\mathbf{r}  = \mathbf{r}_{\text{UL}} + \mathbf{r}_{\text{DL}}+\boldsymbol{w},
\end{equation}
where $\boldsymbol{w} \sim \mathcal{N}(0, \sigma^2 )$ is the additive white Gaussian noise (AWGN) of the channel with the variance $\sigma^2$. Because the uplink and downlink signals often differ in symbol duration and lack temporal synchronization, they inevitably overlap at the receiver. This overlap complicates the separation of two signals.

\section{OFDM Signal in Affine Domain}
In this section, we analyze the behavior of OFDM in the affine domain under coexistence with AFDM signals. The characteristics of the OFDM signal are examined at both the transmitter and receiver. 

\subsection{Pre-Channel Transmission}
Neglecting the CP, by applying an $N$-point DAFT to the time-domain OFDM signal, its corresponding representation in the affine domain can be expressed as
\begin{equation}\label{eqn:ofdm_affine}
\mathbf{s}_\text{A} = \boldsymbol{\Lambda}_{c_1} \mathbf{F}_{N} \boldsymbol{\Lambda}_{c_2}\mathbf{F}^\text{H}_{N} \mathbf{x}_\text{DL},
\end{equation}
As $x_\text{DL}[n]$ are chosen from $\mathcal{M}$-QAM constellation with zero mean and unit variance, when $M$ is large enough $\mathbb{E} \{ \mathbf{x}_\text{DL} \} =0$ and $\mathrm{var} \{ \mathbf{x}_\text{DL} \} =1$. 
As $\boldsymbol{\Lambda}_{c_1}, \mathbf{F}_{N} , \boldsymbol{\Lambda}_{c_2}$ are all unitary matrices,  the covariance of OFDM signal in affine domain $\mathbf{s}_\text{A}$ can be derived as
\begin{align}\label{eqn:covariance}
\mathbf{C}_{s} \!&= \!\mathbb{E}\left\{ \mathbf{\mathbf{s}_\text{A}}\mathbf{\mathbf{s}}_\text{A}^\text{H} \right\} 
\!=\! \mathbb{E}\left\{ \mathbf{\Lambda F}_{N}^\text{H} \mathbf{x}_\text{DL}\mathbf{x}_\text{DL}^\text{H} \mathbf{F}_{N} \mathbf{\Lambda}^\text{H} \right\} \nonumber\\
\!&= \!\mathbf{\Lambda F}_{N}^\text{H} \mathbf{C}_{x} \mathbf{F}_{N} \mathbf{\Lambda}^\text{H}\!= \!\mathbf{I},
\end{align}
where $\mathbf{\Lambda}=\boldsymbol{\Lambda}_{c_1} \mathbf{F}_{N} \boldsymbol{\Lambda}_{c_2}$ and $\mathbf{C}_{x}=\mathbb{E} \left[ \mathbf{x}_\text{DL}\mathbf{x}_\text{DL}^\text{H} \right]=\mathbf{I}$. Hence $\mathbf{C}_{s}$ is a diagonal matrix with entries 1. Therefore, the variance of $\mathbf{s}_\text{A}$ can be calculated by
$\mathrm{var}\{ \mathbf{s}_\text{A} \} =1$, which is a constant value across affine domain.
The expectation of $\mathbf{s}_\text{A}$ can be derived as $\mathbb{E} \{ \mathbf{s}_\text{A} \}= \boldsymbol{\Lambda}\mathbf{F}^\text{H}_{N}\mathbb{E} \{ \mathbf{x}_\text{DL} \}=0$. These findings imply that the OFDM signal exhibits a flat power distribution across the affine domain, resembling the characteristics of AWGN. However, when $N$ is not sufficiently large, the power distribution of the transmitted symbols  $\mathbf{x}_\text{DL}$ is not perfectly uniform. In this case, the covariance matrix of the transmit symbols deviates from the identity matrix, and the resulting $\mathbf{C}_{s}$ is no longer strictly diagonal. Nevertheless, the magnitudes of the off-diagonal components adjacent to the main diagonal remain relatively small, ensuring that the power distribution along the main diagonal of $\mathbf{C}_{s}$ is still approximately uniform. This property can be expressed mathematically as
\begin{equation}
\mathrm{tr}(\mathbf{C}_{s}) \approx \sum_{n=0}^{N-1} \mathbb{E}\left\{ |{x}_\text{DL}[n]|^2 \right\} = N \bar{\mu}, 
\label{eq:trace}
\end{equation}
where $\bar{\mu}$ denotes the average symbol power which is 1 in our case. In particular, the diagonal dominance of $\mathbf{C}_{s}$ becomes more strengthened when $N$ is large. Since $c_1$ and $c_2$ will be small enough, the matrix $\Lambda_{c}$, where $c=c_1, c_2$, can be approximated as an identity matrix. This effectively de-correlates the samples and suppresses the power of the off-diagonal elements.

\subsection{Post-Channel Transmission}
Consider the representation of received OFDM echo signal as $\mathbf{r}_{\text{DL}} = \boldsymbol{\Lambda}\mathbf{\Pi}^\tau \mathbf{\Delta}^{\nu}\mathbf{F}_{N}^\text{H}\mathbf{x}_{\text{DL}},$
where the channel coefficient $h$ is omitted in this expression for simplicity of following derivation. 
To demonstrate that $\mathbf{r}_{\text{DL}}$ exhibits AWGN-like behavior in the affine domain, it is sufficient to prove that the first-order moment of $\mathbf{r}_{\text{DL}}$ remains constant, i.e., $\mathbb{E}\{\mathbf{r}_{\text{DL}}\} = 0$, 
and that the second-order moment depends only on the sample index lag. Specifically,
\begin{equation}
\mathbb{E}\{{r}_{\text{DL}}[n]{r}_{\text{DL}}^{*}[n']\} = R_{rr}[\Delta n], \quad \Delta n = n - n',
\end{equation}
where $R_{rr}[\Delta n]$ denotes the autocorrelation function. 
Moreover, at zero lag, the autocorrelation must satisfy
$R_{rr}[0] = \mathrm{var}\{\mathbf{r}_{\text{DL}}\} = \bar{\mu}$. 
Consider the representation of the received echo OFDM signal in the affine domain as
\begin{equation}
{r}_{\text{DL}}[n] = \sum_{m=0}^{N-1} \sum_{k=0}^{N-1}  x_\text{DL}[k] \, 
e^{j \tfrac{2\pi k}{N}(m-\tau)} 
e^{j \tfrac{2\pi \nu m}{N}} 
\Lambda(n,m),
\end{equation}
Therefore, the autocorrelation function can be derived as
\begin{align}\label{eqn:auto_r}
R_{rr} & \!=\! \mathbb{E}\{{r}_{\text{DL}}[n]{r}_{\text{DL}}^{*}[n']\} \nonumber\\
&\!=\! \sum_{m} \sum_{m'} \Lambda(m,n)\Lambda^{*}(m',n') 
R_{ss}[m-m'] 
e^{j \tfrac{2\pi \nu}{N} (m-m')} \nonumber\\
&\!= \! \sum_{m'} \Lambda^{*}(m',n') 
\sum_{m} R_{ss}[m-m'] \Lambda(m,n) 
e^{j \tfrac{2\pi \nu}{N} (m-m')},
\end{align}
where $R_{ss}[m-m']=\mathbb{E}\!\left\{ s_\text{DL}[m] s_\text{DL}^{*}[m'] \right\} $ is the autocorrelation of the time domain OFDM signal $\mathbf{s}_\text{DL}$. The autocorrelation in affine domain can be obtained by $\sum_{m} R_{ss}[m-m']\, e^{-j 2\pi \left( c_{1} m^{2} + \tfrac{m n}{N} + c_{2} n^{2} \right)}$. 
A shift in time is equivalent to a scaled shift in the affine domain as
\begin{equation}
\Lambda R_{ss}[m-m'] e^{j \frac{2 \pi \nu m}{N}}
\;\xrightleftharpoons[\text{time}]{\text{affine}}\;
R_{ss}^a[n -\theta _{m'}],
\end{equation}
where $R_{ss}^a$ is autocorrelation in affine domain and $\theta _{m'} = \llrrparen{2 N c_{1} m' + \nu }_{N}$ is the localized coupled shift in the affine domain\cite{afdm}, where $\llrrparen{\cdot}_N$ represents the modulo-$N$ operation. Therefore, (\ref{eqn:auto_r}) can be further derived as 
\begin{equation}\label{eqn:auto_r2}
R_{rr}  \!=\! \sum_{M'} R_{ss}^{a}[N - \theta _{m'}] 
\, \Lambda^{*}(m', n') 
\, e^{-j \tfrac{2\pi \nu m'}{N}} .
\end{equation}

Before applying the IDAFT, it is important to note that 
$R_{ss}^{a}[n - \theta _{m'}]$ 
experiences both scaling and reversal with respect to the summation variable $m'$. 
Since $\Lambda$ consists of two diagonal circulant matrices, 
a reversal in the time domain results in a corresponding reversal in the affine domain, 
and vice versa \cite{de2024wide}. 
This relationship can be expressed as  
\begin{equation}
y[n] = \sum_{m} \Lambda(m,n)x[m] 
\;\Longleftrightarrow\;
y[-n] = \sum_{m} \Lambda(m,n)x[-m].
\end{equation}

Furthermore, because the scaling factor $2N c_{1}$ is an integer, 
it determines the affine carriers at $2N c_{1} m'$ 
that perform the IDAFT while preserving periodicity over $N$. 
Consequently, the scaled carriers correspond to time-domain samples scaled by $2N c_{1}$. 
Thus, the autocorrelation function can be rewritten as  
\begin{equation}
R_{rr} = R_{ss}\!\left[ \llrrparen{\Delta n - 2\nu}_{N} \right],
\end{equation}
Therefore, the function depends on the sample lags $\Delta n$. 
Moreover, since $R_{ss}[m] = \bar{\mu}$ is constant in the time domain, 
it follows that 
$R_{yy}[0] = R_{ss}[\llrrparen{-2\nu}_{N}] = \bar{\mu}$, 
is also constant across the affine domain.

For the expectation function 
$\mathbb{E} \left\{\mathbf{r}_{\text{DL}} \right\}=\mathbb{E}\!\left\{ \mathbf{\Lambda} \mathbf{\Pi}^{\tau} \mathbf{\Delta}^{\nu} \mathbf{F}_N^\text{H} \mathbf{x}_\text{DL} \right\}$, 
although the delay and Doppler components are channel-dependent and random, 
they remain independent of the transmit symbols $\mathbf{x}_\text{DL}$. 
Since both $\mathbf{\Pi}^{\tau}$ and $\mathbf{\Delta}^{\nu}$ are linear operations, 
it follows that
\begin{equation}
\mathbb{E}\{\mathbf{r}_{\text{DL}}\} 
= \mathbf{\Lambda} \mathbf{\Pi}^{\tau} \mathbf{\Delta}^{\nu} \mathbf{F}^\text{H}_N \, \mathbb{E}\{\mathbf{x}_\text{DL}\} 
= 0.
\label{eqn:expect_rs}
\end{equation}

\begin{figure}
  \centering
    \begin{minipage}[t]{0.235\textwidth}
    \includegraphics[width=\textwidth]{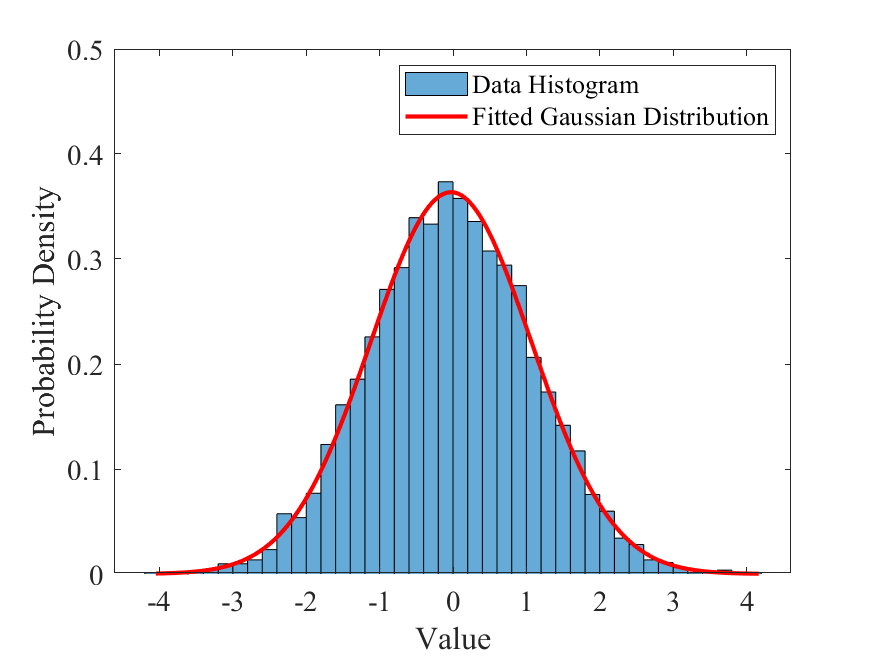}
    \centering{\footnotesize  {(a) Real part }}
    \label{fig:sub3}
  \end{minipage}
  \begin{minipage}[t]{0.235\textwidth}
    \includegraphics[width=\textwidth]{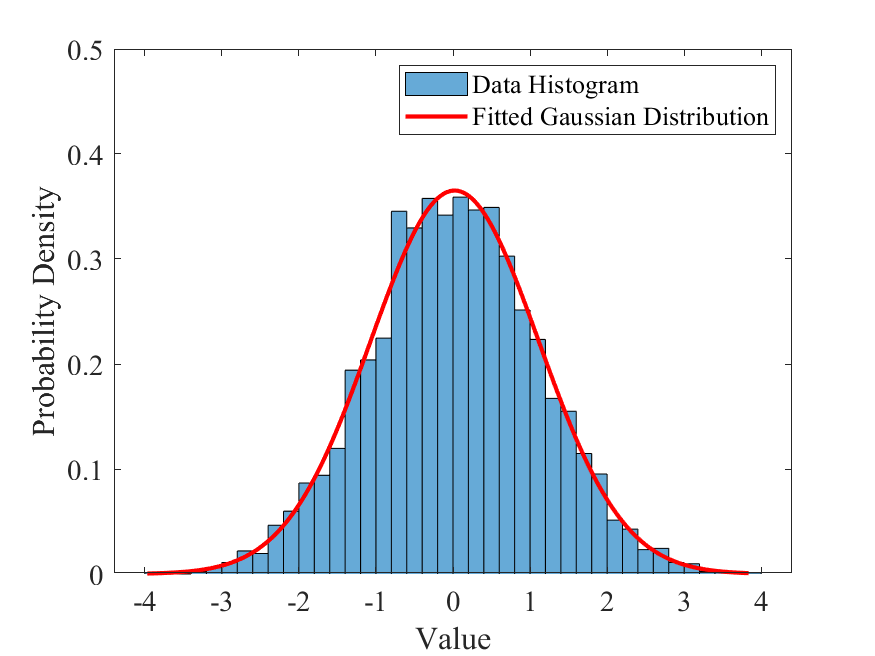}
    \centering{\footnotesize {(b) Imaginary part }}
    \label{fig:sub4}
  \end{minipage}
  \vspace{-0.1cm}
  \caption{Histogram of post- channel OFDM signal in affine domain with fitted Gaussian distribution.}
  \vspace{-0.3cm}
  \label{fig:compare}
\end{figure}

Therefore, the OFDM signal exhibits constant power and can therefore be modeled as AWGN-like in the affine domain. Consequently, the combined effect of $\mathbf{r}_\text{DL}$ and $\boldsymbol{w}$ in (\ref{eqn:r}) can be interpreted as an effective superposition of noise. Fig.\ref{fig:compare} shows the histogram of the OFDM signal in the affine domain at the receiver. It can be observed that the OFDM samples exhibit an AWGN-like distribution, which supports and validates the analytical proof. 

\section{AFDM Frame Design for Noise Power Estimation and Data Detection}
For data detection, the BS decodes the communication signal while modeling both the echo signal and noise as interference. 
A central limitation of this approach is that: the echo signal interference is typically colored and non-Gaussian for PD-NOMA ISAC system. Modeling it as white Gaussian noise produces a mismatched likelihood, which elevates the decoding error probability. The resulting detection errors leave residual interference after cancellation, which propagates into the second stage and degrades the targets parameters estimation performance. 

In contrast, this issue does not arise in the WD-NOMA ISAC system, since the OFDM echo signal exhibits AWGN-like behavior in the affine domain and can therefore be treated as white noise. Assuming that the uplink channel is known at the receiver, the AFDM symbols can be reliably detected using an MMSE detector. Nevertheless, as the effective noise in the received signal originates from both the  channel noise and the echo component, the total noise power must be estimated to enable accurate MMSE detection. To address this, we design a dedicated AFDM frame structure that facilitates noise power estimation (NPE).

\begin{figure}[t]
\centering
\scalebox{0.55}{
\begin{tikzpicture}[font=\sffamily]

\def\nboxes{21} 
\def\boxsize{0.6}
\def\dataStart{6} 
\def\dataEnd{15}  
\def\leakEnd{9} 
\def\leakStart{13}  
\def\leakEndd{16} 
\def\leakStartt{12}  
\definecolor{pilot}{RGB}{255,220,120}
\definecolor{guard}{RGB}{255,245,210}
\definecolor{data}{RGB}{255,170,100}

\foreach \i in {1,...,\nboxes}{
    \ifnum\i=19
        \node[draw,fill=guard,minimum size=\boxsize cm] (bs\i) at (\i*\boxsize,0) {...};
    \else\ifnum\i=3
        \node[draw,fill=guard,minimum size=\boxsize cm] (bs\i) at (\i*\boxsize,0) {...};
    \else\ifnum\i=11
        \node[draw,fill=data,minimum size=\boxsize cm] (bs\i) at (\i*\boxsize,0) {...};    
    \else\ifnum\i<3
        \node[draw,fill=guard,minimum size=\boxsize cm] (bs\i) at (\i*\boxsize,0) {$d$};
    \else\ifnum\i<7
        \node[draw,fill=guard,minimum size=\boxsize cm] (bs\i) at (\i*\boxsize,0) {$d$};
    \else\ifnum\i<11
        \node[draw,fill=data,minimum size=\boxsize cm] (bs\i) at (\i*\boxsize,0) {0};
    \else\ifnum\i<17
        \node[draw,fill=data,minimum size=\boxsize cm] (bs\i) at (\i*\boxsize,0) {0};
    \else\ifnum\i<19
        \node[draw,fill=guard,minimum size=\boxsize cm] (bs\i) at (\i*\boxsize,0) {$d$};    
    \else
        \node[draw,fill=guard,minimum size=\boxsize cm] (bs\i) at (\i*\boxsize,0) {$d$};
    \fi\fi\fi\fi\fi\fi\fi\fi
}

\node[anchor=west] at (bs\nboxes.east) {transmitter};


\draw[decorate,decoration={brace,amplitude=5pt}] (bs1.north west) -- (bs6.north east) node[midway,yshift=14pt]{data};
\draw[decorate,decoration={brace,amplitude=5pt}] (bs10.north west) -- (bs12.north east) node[midway,yshift=14pt]{Guard for NPE};
\draw[decorate,decoration={brace,amplitude=5pt}] (bs17.north west) -- (bs21.north east) node[midway,yshift=14pt]{data};
\draw[decorate,decoration={brace,mirror,amplitude=5pt}] (bs7.south west) -- (bs9.south east) node[midway,yshift=14pt]{};
\draw[decorate,decoration={brace,mirror,amplitude=5pt}] (bs13.south west) -- (bs16.south east) node[midway,yshift=14pt]{};
\node[anchor=west] at (5.1,-0.8) {Guard for channel};
\draw[->] (4.5,-0.6) -- (5.1,-0.8) ;
\draw[->] (8.75,-0.6) -- (8.15,-0.8) ;
\foreach \i in {1,...,\nboxes}{
    \ifnum\i=4
        \node[draw,fill=pilot,minimum size=\boxsize cm] (ue\i) at (\i*\boxsize,-1.5) {...};
    \else\ifnum\i=11
        \node[draw,fill=data,minimum size=\boxsize cm] (ue\i) at (\i*\boxsize,-1.5) {...};
    \else\ifnum\i=18
        \node[draw,fill=pilot,minimum size=\boxsize cm] (ue\i) at (\i*\boxsize,-1.5) {...};
    \else\ifnum\i<4
        \node[draw,fill=pilot,minimum size=\boxsize cm] (ue\i) at (\i*\boxsize,-1.5) {$\tilde{d}$};
    \else\ifnum\i<10
        \node[draw,fill=pilot,minimum size=\boxsize cm] (ue\i) at (\i*\boxsize,-1.5) {$\tilde{d}$};
    \else\ifnum\i<11
        \node[draw,fill=data,minimum size=\boxsize cm] (ue\i) at (\i*\boxsize,-1.5) {0};
    \else\ifnum\i<13
        \node[draw,fill=data,minimum size=\boxsize cm] (ue\i) at (\i*\boxsize,-1.5) {0};
    \else\ifnum\i<19
        \node[draw,fill=pilot,minimum size=\boxsize cm] (ue\i) at (\i*\boxsize,-1.5) {$\tilde{d}$};    
    \else
        \node[draw,fill=pilot,minimum size=\boxsize cm] (ue\i) at (\i*\boxsize,-1.5) {$\tilde{d}$};
    \fi\fi\fi\fi\fi\fi\fi\fi
}

\node[anchor=west] at (ue\nboxes.east) {receiver};

\draw[decorate,decoration={brace,mirror,amplitude=5pt}] (ue7.south west) -- (ue9.south east) node[midway,yshift=-12pt]{$\kappa_\text{max}$};
\draw[decorate,decoration={brace,mirror, amplitude=5pt}] (ue13.south west) -- (ue16.south east) node[midway,yshift=-12pt]{$\kappa_\text{max}+2Nc_1n'$};


\end{tikzpicture}
}
\vspace{-0.5cm}
\caption{Proposed AFDM frame design.}
\vspace{-0.6cm}
\label{fig:guard}
\end{figure}
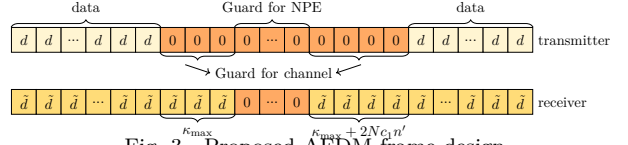
The design of the AFDM frame structure is illustrated in Fig. \ref{fig:guard}. In this structure, each AFDM symbol is allocated $K$ consecutive guard subcarriers, which are used for NPE. $d$ denotes the transmitted QAM symbols at the transmitter, while $\tilde{d}$ represents the data symbols obtained after CPP removal and the DAFT operation at the receiver. The parameter $\kappa_\text{max}$ denotes the maximum normalized integer Doppler shift. Each data symbol spreads over $\kappa_\text{max}$ subcarriers on the right and $\kappa_\text{max}+2Nc_1n'$ on the left at the receiver after demodulation\cite{afdmisac}.
The AFDM subcarrier grid is partitioned into data and two disjoint guard subsets:
(i) a \emph{channel guard} $\mathcal{G}_1$ sized to absorb delay/Doppler leakage, and
(ii) an \emph{NPE guard} $\mathcal{G}_2$ reserved for channel noise and OFDM interference power estimation.
The allocation is
\begin{equation}
x_{\mathrm{a}}[k]=
\begin{cases}
0,      & k \in \mathcal{G}_1 \quad \text{(channel guard)},\\[2pt]
0,      & k \in \mathcal{G}_2 \quad \text{(NPE guard)},\\[2pt]
d[k],   & k \in \mathcal{D} \quad \text{(data region)},\\
\end{cases}
\label{eq:alloc}
\end{equation}
where $\mathcal{G}=\mathcal{G}_1 \cup \mathcal{G}_2$, $\mathcal{G}_1 \cap \mathcal{G}_2=\varnothing$,
and $\mathcal{D}=\{0,1,\ldots,N{-}1\}\setminus\mathcal{G}$.
The guard subsets are contiguous blocks on an $N$-point grid
$\mathcal{G}_1 = \{\,i,\,i{+}1,\,\ldots,\,i{+}K_1{-}1\,\},  
\mathcal{G}_2 = \{\,j,\,j{+}1,\,\ldots,\,j{+}K_2{-}1\,\},$
with total guard length $K=K_1{+}K_2$ kept fixed across frames.
\noindent
 $K_1$ is chosen to cover the anticipated integer dispersion of delay and Doppler, while $K_2$ sets the number of i.i.d. samples available for the variance estimator. Indices $i$ and $j$ may be placed contiguously, wrap-around is handled by the modulo-$N$ convention above.

Under the assumed LTV channel model with both integer delay and integer Doppler, ICI is restricted to the subcarriers adjacent to the data subcarriers. When the guard region is sufficiently long, the central guard subcarriers remain free from ICI, as illustrated in Fig.~\ref{fig:guard}. These unaffected subcarriers therefore contain only the echo signal and channel noise components. 
The symbols after demodulation can be represented as 
\begin{equation}
\mathbf{\tilde{d}} =  \boldsymbol{\Lambda}_{c_2}\mathbf{F}_{N} \boldsymbol{\Lambda}_{c_1}\mathbf{B}_\text{cpp}\mathbf{r}, 
\end{equation}
where $\mathbf{B}_\text{cpp}=[\mathbf{0}_{N \times {L_{\rm cpp}}}, \mathbf{I}_{N}]$ is the CPP removal matrix. 
The total power can be estimated from the uncontaminated subcarriers as $
\widehat{\sigma}^2 = \mathrm{var} \{ \tilde{d}[k]\}, i+\kappa_\text{max}<k<i+K-\kappa_\text{max}-2Nc_1n$. 
The equivalent channel of AFDM signal can be written as 
\begin{equation}
\mathbf{H}_\text{a}^\text{eq} = \boldsymbol{\Lambda}_{c_2}\mathbf{F}_{N} \boldsymbol{\Lambda}_{c_1}\mathbf{B}_\text{cpp}\mathbf{H}_\text{c}\mathbf{A}_\text{cpp}\boldsymbol{\Lambda}_{c_1}^\text{H} \mathbf{F}^\text{H}_{N} \boldsymbol{\Lambda}_{c_2}^\text{H} 
\end{equation}
Finally, the symbol detection can be performed using MMSE as 
\begin{equation}
\mathbf{\widehat{x}_\text{UL}} = \Big( (\mathbf{H}_\text{a}^\text{eq})^{\rm H} \mathbf{H}_\text{a}^\text{eq} + \widehat{\sigma}^2 \mathbf{I}_{N} \Big)^{-1} (\mathbf{H}_\text{a}^\text{eq})^{\rm H} {{\mathbf{\tilde{d}}}}.
\end{equation} 

\section{OMP Based Targets Detection for Sensing}
After recovering the uplink transmit symbols, the echo component together with the noise can be obtained by subtracting the reconstructed AFDM signal from the received signal. The post-cancellation signal is written as
\begin{equation}
\widehat{\mathbf{r}}_{\text{DL}}=    \mathbf{r} - \mathbf{H}_\text{UL}\mathbf{W}_\text{AFDM}\widehat{\mathbf{x}}_{\text{UL}}
\end{equation}

In this section, a 2D-OMP algorithm is employed to iteratively estimate and remove dominant target components from the post-cancellation signal. 
In this algorithm, a sensing dictionary $\mathbf{\Phi}=[\mathbf{\phi}_{0,0}, \mathbf{\phi}_{0,1}, \ldots, \mathbf{\phi}_{N_\tau-1,N_\nu-1}]$ is pre-generated before algorithm implementation,  where $N_\tau$ and $N_\nu$ are the grid size of delay and Doppler respectively. Each entry of the dictionary is constructed by multiplying the transmit downlink OFDM signal with a phase term corresponding to a specific Doppler shift and applying a certain delay. Therefore, it produces a delayed and Doppler-shifted replica of the OFDM signal. Each dictionary entry can be expressed as
$\mathbf{\phi}_{ij} \!=\!
\left[
x_s[0-\tau_i] e^{\frac{j 2 \pi \nu_j 0}{N}}, \;
\ldots, \;
x[N-1-\tau_i] e^{\frac{j 2 \pi \nu_j (N-1)}{N}}
\right]^{T}.$

At each iteration, a two-dimensional grid search is performed to identify the delay–Doppler pair that yields the maximum correlation between the post-cancellation signal and the transmitted DL OFDM signal. The corresponding component associated with the detected target is then reconstructed and subtracted from the post-cancellation signal. This procedure is repeated until all targets are estimated and removed, leaving a residual signal that consists of noise only. The overall procedure is summarized in Algorithm 1. 

\begin{algorithm}[t]
\caption{Unified WD–NOMA Receiver: NPE, MMSE, Cancellation, 2D–OMP}
\begin{algorithmic}[1]
\State \textbf{Inputs:} $\mathbf{r}$, $\mathbf{B}_{\rm cpp},\mathbf{\Lambda}_{c_1},\mathbf{\Lambda}_{c_2},\mathbf{F}_N$, $\mathbf{H}_\text{c} $, $\mathbf{\Phi}$, $\mathcal{G}_2$, $\kappa_{\max}$, $P$
\State \textbf{Outputs:} $\hat{\mathbf{x}}_a$, $\mathcal{T}$

\medskip
\State \textbf{Phase A (NPE)}
\State $\widetilde{\mathbf d} \gets \boldsymbol{\Lambda}_{c_2}\mathbf{F}_{N} \boldsymbol{\Lambda}_{c_1}\mathbf{B}_\text{cpp}\mathbf{r}$
\State $\mathcal{I}_{\rm NPE} \gets \{\,k\in\mathcal{G}_2 \mid i{+}\kappa_{\max} < k < i{+}K{-}\kappa_{\max}-2Nc_1n\,\}$
\State $\widehat{\sigma}^2 \gets \frac{1}{|\mathcal{I}_{\rm NPE}|}\!\sum_{k\in\mathcal{I}_{\rm NPE}} |\widetilde d[k]|^2$ \hfill 

\medskip
\State \textbf{Phase B (MMSE)}
\State $\mathbf{H}_\text{a}^\text{eq} \gets \boldsymbol{\Lambda}_{c_2}\mathbf{F}_{N} \boldsymbol{\Lambda}_{c_1}\mathbf{B}_\text{cpp}\mathbf{H}_\text{c}\mathbf{A}_\text{cpp}\boldsymbol{\Lambda}_{c_1}^\text{H} \mathbf{F}^\text{H}_{N} \boldsymbol{\Lambda}_{c_2}^\text{H}$
\State $\mathbf{\widehat{x}_\text{a}} \gets \Big( (\mathbf{H}_\text{a}^\text{eq})^{\rm H} \mathbf{H}_\text{a}^\text{eq} + \widehat{\sigma}^2 \mathbf{I}_{N} \Big)^{-1} (\mathbf{H}_\text{a}^\text{eq})^{\rm H} {{\mathbf{\tilde{d}}}}$

\medskip
\State \textbf{Phase C (Cancellation)}
\State $\widehat{\mathbf{s}}_\text{a} \gets \mathbf{H}_a \mathbf{A}_\text{cpp}\boldsymbol{\Lambda}_{c_1}^\text{H} \mathbf{F}^\text{H}_{N} \boldsymbol{\Lambda}_{c_2}^\text{H}\mathbf{\widehat{x}_\text{a}} $
\State $\widehat{\mathbf r}_\text{o} \gets \mathbf{r} - \widehat{\mathbf{s}}_\text{a}$

\medskip
\State \textbf{Phase D (2D–OMP)}
\State $\mathbf{r}' \gets \widehat{\mathbf r}_\text{o}$, \quad $\mathcal{T}\gets\varnothing$
\For{$p=0$ \textbf{to} $P-1$}
    \State $\mathbf{R} \in \mathbb{C}^{N_\tau \times N_\nu}$
    \For{$i=0$ \textbf{to} $N_\tau-1$}
        \For{$j=0$ \textbf{to} $N_\nu-1$}
            \State $\mathbf{R}[i,j] \gets \big|\boldsymbol{\phi}_{ij}^{H}\mathbf{r}'\big|$
        \EndFor
    \EndFor
    \State $(i^\star,j^\star) \gets \arg\max_{i,j}\, \mathbf{R}[i,j]$
    \State $\hat{\tau}_p \gets \tau_{i^\star}$, \; $\hat{\nu}_p \gets \nu_{j^\star}$, \;
           $\hat{h}_p \gets (\boldsymbol{\phi}_{i^\star j^\star})^{\dagger}\mathbf{r}'$
    \State $\mathcal{T} \gets \mathcal{T} \cup \{(\hat{\tau}_p,\hat{\nu}_p)\}$
    \State $\mathbf{r}' \gets \mathbf{r}' - \hat{h}_p \boldsymbol{\phi}_{i^\star j^\star}$
\EndFor
\State \Return $\hat{\mathbf{x}}_a, \mathcal{T}$
\end{algorithmic}
\end{algorithm}

\section{Numerical Results}
In this section, we present simulation results to evaluate the performance of the proposed algorithm. The simulation parameters are summarized in Table \ref{tab1} .
\begin{table}
\centering
\caption{Simulation Parameters}
\resizebox{0.3\textwidth}{!}
{\begin{tabular}{| c || c |}
\hline\hline
\textbf{Simulation Parameter}
& \textbf{Value}
\\ \hline \hline
Subcarriers
& $N=1024$
\\ \hline
QAM size
& $\mathcal{M}=4$
\\ \hline
Carrier frequency
& $f_c=28~\rm{GHz}$
\\ \hline
subcarrier spacing
& $B=30~\rm{kHz}$
\\ \hline
Guard index
& $0, 1, \ldots, 63 $
\\ \hline
speed
& $v=500~\rm{km/h}$

\\ \hline
Number of targets 
& $2$
\\ \hline
Target range 
& $[0,50m] $
\\ \hline
Target velocity 
& $[0,500~\rm{km/h}] $
\\ \hline\hline
\end{tabular}
\label{tab1}
}
\vspace{-0.5cm}
\end{table}

\begin{figure}[!t]
\centering
\includegraphics[scale=0.4]{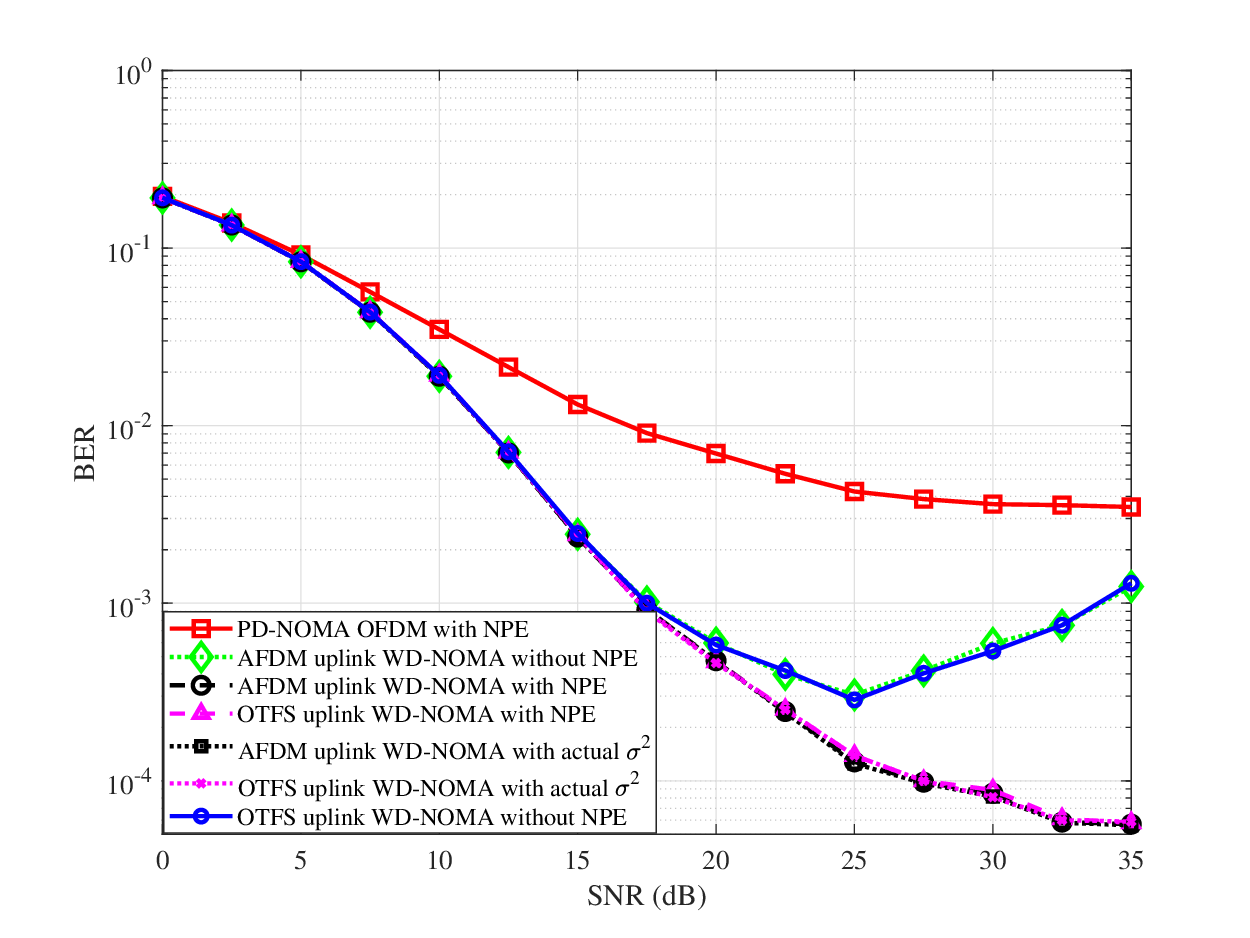} 
\vspace{-0.3cm}
\caption{BER performance of different ISAC systems.}
\label{fig:ber}
\vspace{-0.5cm}
\end{figure}

Fig. \ref{fig:ber} illustrates the BER performance of the uplink communication signal over a three-tap Rayleigh fading channel of integer Doppler shift. MMSE detector is employed for data symbol detection. Different ISAC system are compared as shown in the figure. The PD-NOMA system using OFDM for uplink has the worst BER performance because OFDM behaves bad in fast time-varying channel compare to AFDM. For all WD-NOMA ISAC systems, the BER performance are all identical when the SNR is below 15 dB. However, for WD-NOMA ISAC system without NPE, the BER performance degradation is observed once the SNR exceeds 20 dB. This occurs because the sensing signal is set 20 dB weaker than the communication signal, as the SNR surpasses 20 dB, the sensing interference becomes the dominant noise component. Consequently, the actual effective SNR is significantly higher than the SNR determined solely by the channel noise. Therefore, without NPE, the MMSE detector relies on an inaccurate noise power, which leads to considerable estimation errors. In contrast, with the proposed NPE, the BER continues to decrease as the SNR increases beyond 20 dB and it has the same results as actual total noise power is known. This demonstrates that the effectiveness of the proposed NPE method. In addition, we also simulate the WD-NOMA ISAC system using OTFS as uplink signal. In this case, the Doppler bins on the two sides of OTFS frame are used for NPE. The results indicate no significant difference in BER performance when using AFDM or OTFS as the uplink waveform in the WD-NOMA system. This highlights that both DFT-s-OFDM-based waveforms offer notable BER performance improvements compared with OFDM.

\begin{figure}[!t]
\centering
\includegraphics[scale=0.4]{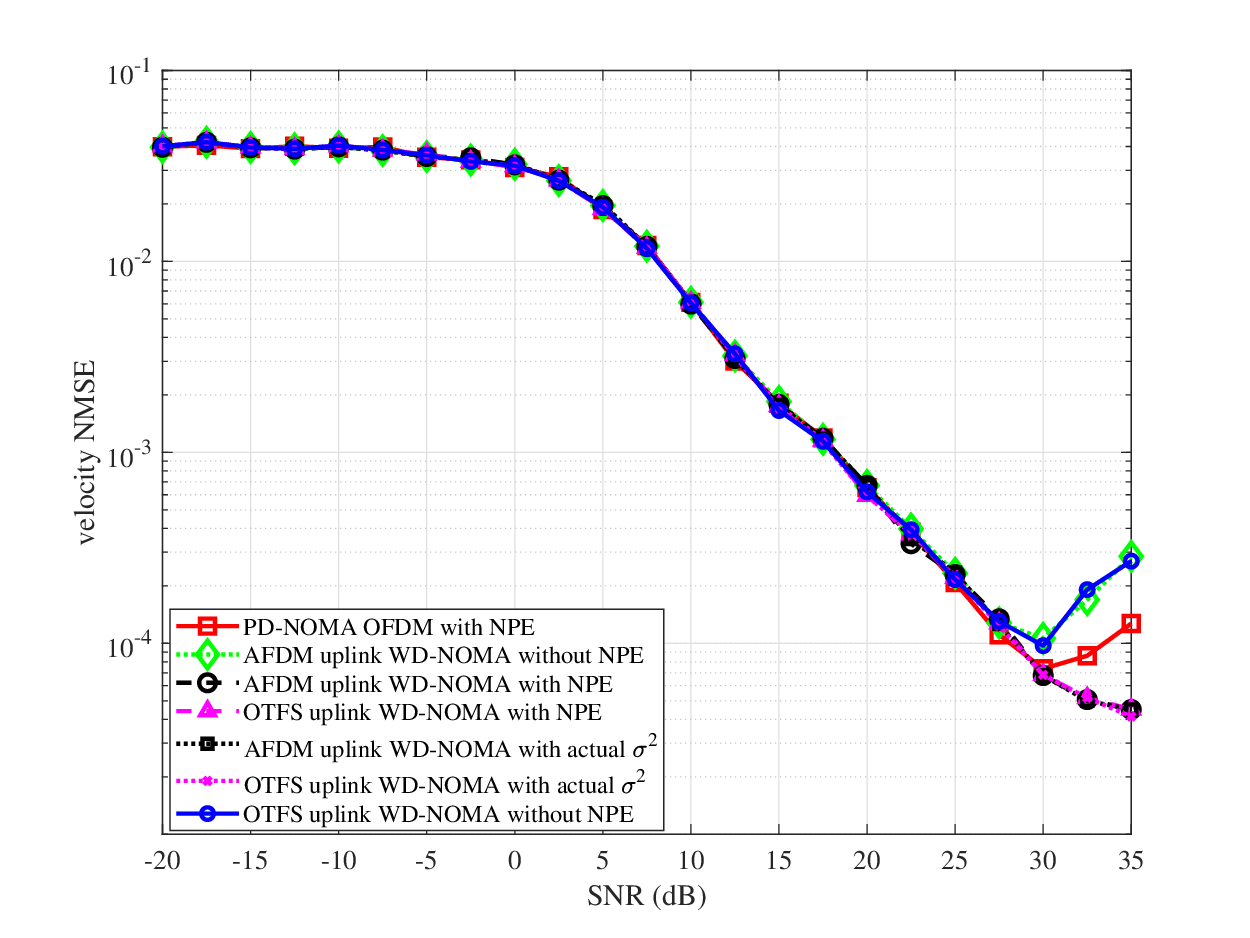} 
\vspace{-0.3cm}
\caption{Velocity estimation performance of different ISAC systems.}
\label{fig:distance}
\vspace{-0.5cm}
\end{figure}

\begin{figure}[!t]
\centering
\includegraphics[scale=0.4]{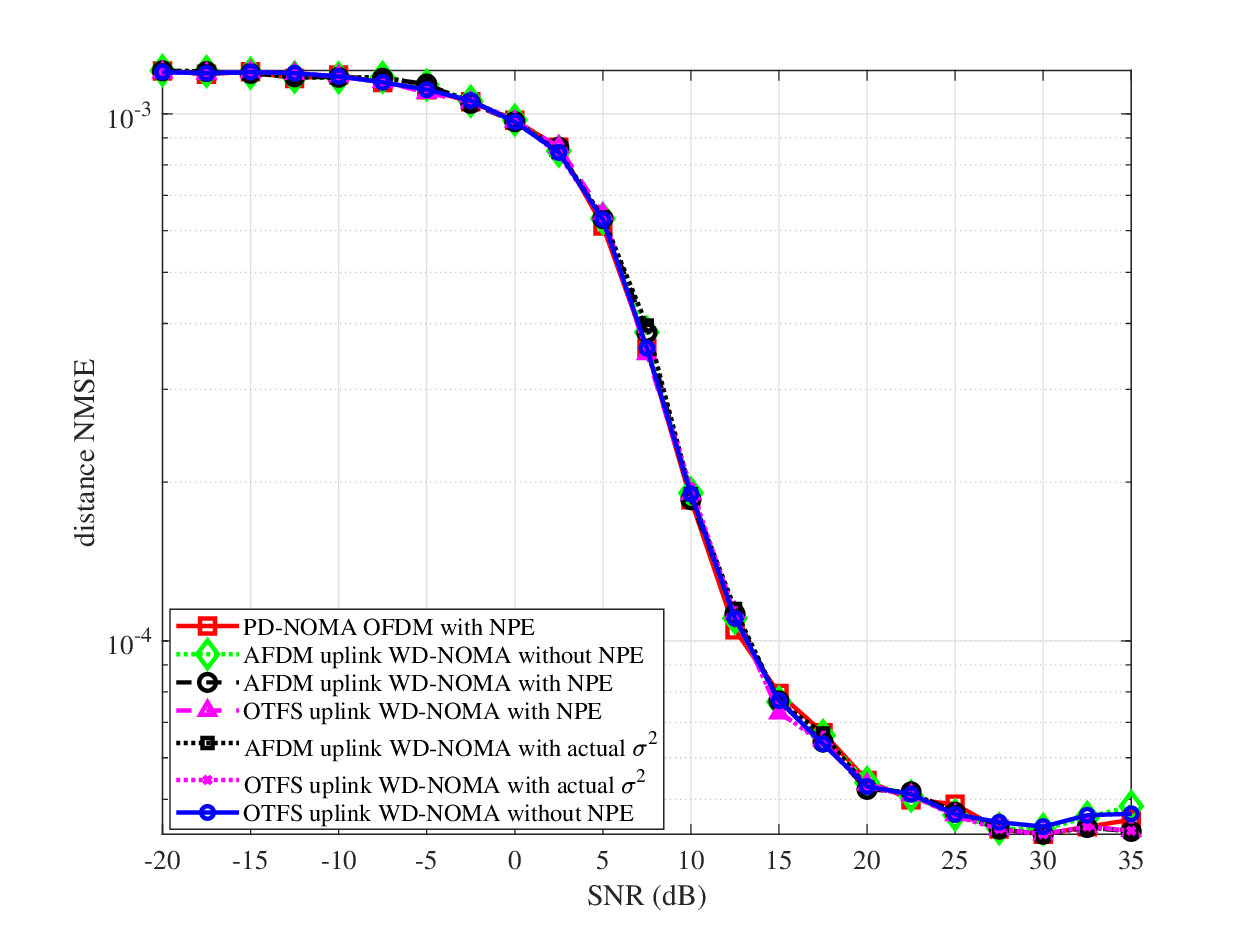} 
\vspace{-0.3cm}
\caption{Distance estimation performance of different ISAC systems.}
\label{fig:vel}
\vspace{-0.5cm}
\end{figure}

Figs. \ref{fig:distance} and \ref{fig:vel} present the normalized mean square error (NMSE) of the velocity and distance estimations for all targets. For velocity estimation, the NMSE remains nearly identical across all evaluated systems, with the WD-NOMA system incorporating NPE beginning to exhibit noticeable accuracy improvements beyond 30 dB. For distance estimation, no significant difference in accuracy is observed among the systems. This is attributed to the low distance resolution in the simulation setup and the high precision of the 2D-OMP method, which yields negligible delay estimation errors.

\section{Conclusion}
This paper presented a WD-NOMA framework for ISAC in alignment with recent 6G air-interface developments, where DFT-s-OFDM based waveform such as AFDM and OTFS is selected for uplink and OFDM for downlink transmission. The study primarily focused on the AFDM-based implementation. It was shown that the OFDM sensing signal exhibits AWGN-like characteristics in the affine domain, enabling it to be modeled as white noise during uplink communication signal detection. To ensure accurate data recovery, an AFDM frame design and an NPE method were developed, and a 2D-OMP algorithm was applied for joint distance and velocity estimation of sensing targets. Simulation results verified that the WD-NOMA ISAC system, employing either AFDM or OTFS as UL signal, outperforms the PD-NOMA ISAC system based solely on the OFDM waveform in terms of BER performance. Moreover, the proposed NPE method further enhances BER accuracy.
\bibliographystyle{IEEEtran} 
\bibliography{reference}

\end{document}